\pdfoutput=1

\documentclass[%
 reprint,
 amsmath,amssymb,
 aps,
pra,
longbibliography,
]{revtex4-1}

\usepackage[T1]{fontenc}
\usepackage[latin9]{inputenc}
\usepackage{textcomp}
\usepackage{mathtools}
\usepackage{amsmath}
\usepackage{amssymb}
\usepackage{graphicx}

\makeatletter

\pdfpageheight\paperheight
\pdfpagewidth\paperwidth

\makeatother

\usepackage{babel}

\begin{document}
\title{Error correction for encoded quantum annealing revisited}
\author{Yoshihiro Nambu}
\affiliation{NEC-AIST Quantum Technology Cooperative Research Laboratory~\\
 National Institute of Advanced Industrial Science and Technology }
\begin{abstract}
F. Pastawski and J. Preskill \cite{Pastawski2016} discussed error correction of quantum annealing (QA) based on a parity-encoded spin system, known as the Sourlas-Lechner-Hauke-Zoller (SLHZ) system \cite{Sourlas2005,Lechner2015}. 
They pointed out that the SLHZ system is closely related to a classical low-density parity-check (LDPC) code and demonstrated its error-correcting capability through a belief propagation (BP) algorithm assuming independent random spin-flip errors. 
In contrast, Albash et al. suggested that the SLHZ system does not receive the benefits of post-readout decoding \cite{Albash2016}. 
The reason is that independent random spin-flips are not the most relevant error arising from sampling excited states during the annealing process, whether in closed or open system cases. 
In this work, we revisit this issue: we propose a very simple decoding algorithm to eliminate errors in the readout of SLHZ systems and show experimental evidence suggesting that SLHZ system exhibits error-correcting capability in decoding annealing readouts. 
Our new algorithm can be thought of as a bit-flipping algorithm for LDPC codes. 
Assuming an independent and identical noise model, we found that the performance of our algorithm is comparable to that of the BP algorithm. 
The error correcting-capability for the sampled readouts was investigated using Monte Carlo calculations that simulate the final time distribution of QA.
The results show that the algorithm successfully eliminates errors in the sampled readouts under conditions where error-free state or even code state is not sampled at all. 
Our simulation suggests that decoding of annealing readouts will be successful if the correctable states can be sampled by annealing, and annealing can be considered to play a role as a pre-process of the classical decoding process. 
This knowledge will be useful for designing and developing practical QA based on the SLHZ system in the near future.
\end{abstract}
\maketitle

\section{Introduction}

In 2016, F. Pastawski and J. Preskill (PP) published the paper entitled ``Error correction for encoded quantum annealing''\cite{Pastawski2016}.
They considered the parity-encoded spin system, which first appeared in a paper written by Sourlas in 2005 \cite{Sourlas2005} and rediscovered by W. Lechner, P. Hauke, and P. Zoller in 2015  \cite{Lechner2015}.
They noted that this system, which we call the SLHZ system, is closely related to a classical LDPC code and analyzed its error-correcting capability by belief propagation (BP) algorithm \cite{Pastawski2016}, which is known to be an efficient decoding algorithm for the LDPC codes. 
At about the same time, T. Albash, W. Vincl, and D. Lidar reported the performance of the SLHZ system using simulated quantum annealing (SQA) based on the classical Monte Carlo method. 
They compared the performance of the SLHZ system and that of the spin system arranged according to the minor embedding (ME) scheme \cite{Albash2016}. 
They found that the ME system outperformed the SLHZ system when identical simulation parameters were used and questioned the SLHZ system's error-correcting capability. 
They speculated that while the SLHZ system can correct errors arising from weakly correlated spin-flip noise, it can not correct thermal and dynamical errors in the readouts arising from the excited states sampled during annealing processes in both the closed and open system cases.

In this paper, we revisit this issue and answer the following question: Can the error-correcting codes embedded in the SLHZ system contribute to solving optimization problems? 
Our answer is affirmative. We show evidence that the SLHZ system has error-correcting capability for thermal errors and can eliminate errors even those originating from correlated spin-flip noise. 
In Sec.\ref{sec:2}, we introduced a simple decoding algorithm for the readouts of the SLHZ system to eliminate errors in them. 
Our algorithm was based on majority logic decoding, which was discussed in our previous paper \cite{Nambu2024}. 
In this paper, we compare its performance with that of the BP algorithm given by PP.
Our algorithm has two features. 
First, it uses weight-3 bipolar syndromes to eliminate errors. 
Second, each physical spin orientation is decided by an iterative calculation. 
These features are crucial for decoding the readout of the SLHZ system, which embodies an LDPC code, because it is nothing but a bit-flipping (BF) decoding algorithm for the LDPC codes. 
In Sec.\ref{sec:3}, we investigated the error-correcting capability of our algorithm using the readouts sampled by rejection-free Markov Chain Monte Carlo (RFMCMC) simulation.
This would be reasonable to some extent.  
It is recognized that the characteristics of the QA device reflect its final-state distribution, not its dynamics \cite{Amin2015}. 
Moreover, several studies suggest that currently available QA may be suitable for a fast thermal sampler \cite{Marshall2017,Sarandy2005,Sarovar2013,Amin2015,Benedetti2016}. 
RFMCMC is a very efficient method to sample states from the Maxwell-Boltzmann (MB) distribution excluding state multiplicity due to self-loop transitions \cite{Rosenthal2021,Nambu2024}. 
This suggests that if our decoding algorithm is effective for the readouts sampled by RFMCMC simulations, it is also effective for those sampled by actual QA. 
It was proved that our decoding algorithm can efficiently eliminate errors in the readout of RFMCMC simulation by tuning the hyperparameters assumed in the Hamiltonian of the SLHZ system.  
The present results also imply the correctness of PP's insight that hybrid computation using classical decoding and QA together can solve problems that cannot be solved by either one alone.

\section{Majority logic decision based decoding algorithm\label{sec:2}}
\begin{figure}[tb]
\begin{centering}
\includegraphics[viewport=320bp 200bp 580bp 380bp,clip,scale=0.92]{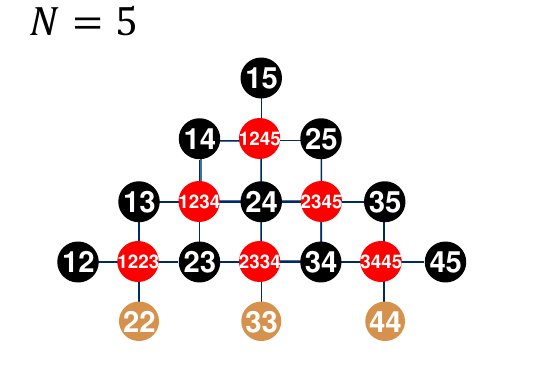}
\caption{
SLHZ system with $N=5$ logical spins. 
Dark blue circles indicate physical spins and yellow-brown circles indicate auxiliary spins with a fixed orientation. 
The label $\left\{ i,j\right\} $ in the circle specifies a physical spin $z_{ij}$. 
The red circle with the label $\left\{ i,j,k,l\right\} $ designates weight-4 syndrome $S_{ijkl}^{(4)}\left(\boldsymbol{z}\right).
$\label{fig1}}
\end{centering}
\end{figure}
Let me start by presenting our decoding algorithm. 
Fig.\ref{fig1} is a graph showing how the spins in the SLHZ system are interconnected for $N=5$ logical spins. 
In this graph, the dark blue circles labeled by a set of two distinct positive integers $\left\{ i,j\right\}$ $(1\leq i<j\leq N)$ represent $\binom{N}{2}$ physical spins, each of which takes the values $z_{ij}\left(=\pm1\right)$ depending on its orientation. 
The yellow-brown circles labeled by $\left\{ i,i\right\}$ represent ancillary spins fixed at positive orientation: $z_{ii}=+1$. 
Every variable $z_{ij}$ encodes the product of logical spin variables $Z_{i}Z_{j}\left(=\pm1\right)$, which is called a code state. 
On the other hand, the red circles labeled by the set $\left\{ i,j,k,l\right\}$ $(1\leq i<j\leq k<l\leq N)$ represent weight-4 syndromes $S_{ijkl}^{(4)}\left(\boldsymbol{z}\right)=z_{ik}z_{jk}z_{jl}z_{il}\left(=\pm1\right)$ (stabilizers for quantum case), 
where we introduced an $N\times N$ symmetric bipolar matrix $\boldsymbol{z}$ whose entries are given by $z_{ij}$ for convenience. 

It is important to note that instead of weight-4 syndromes $S_{ijkl}^{(4)}\left(\boldsymbol{z}\right)$, weight-3 syndromes $S_{ijk}^{(3)}\left(\boldsymbol{z}\right)=z_{ij}z_{jk}z_{ik}\left(=\pm1\right)$ are chosen for any possible set $\left\{ i,j,k\right\} $ $(1\leq i<j<k\leq N)$. 
This is because both $S_{ijk}^{(3)}\left(\boldsymbol{z}\right)=+1$ and $S_{ijkl}^{(4)}\left(\boldsymbol{z}\right)=+1$ for possible  $\left\{ i,j,k,l\right\}$ provide equivalent parity check equations that preserve $z_{ij}=Z_{i}Z_{j}$. 
Although we can consider $\binom{N}{3}$ weight-3 syndromes, only $\binom{N-1}{2}$ of them are independent since there are associated $\binom{N-1}{2}$ independent parity check equations. 
Any given syndrome can be written as a product of the other chosen independent syndromes. 
We note that $\binom{N-1}{2}$ elements of weight-4 syndromes shown in Fig.\ref{fig1} are independent. 
Thus, any $\binom{N}{3}$ weight-3 syndromes $S_{ijk}^{(3)}$ can be deduced from the product of the appropriately chosen weight-4 syndrome $S_{ijkl}^{(4)}$ in Fig.\ref{fig1}. 
For example, for any code state $\boldsymbol{z}$, $S_{135}^{(3)}=S_{1234}^{(4)}S_{1245}^{(4)}S_{2345}^{(4)}S_{2334}^{(4)}$, $S_{145}^{(3)}=S_{1245}^{(4)}S_{2345}^{(4)}S_{3445}^{(4)},$ and so on. 
Consider a set of $N-1$ physical spin variables and a set of the $\binom{N-1}{2}$ weight-3 syndromes with a common index $i$, i.e.,
\begin{subequations}
\label{eq:1}
\begin{equation}
\boldsymbol{z}_{i}=\left\{ z_{ij},z_{ik},z_{il},z_{im},\cdots\:\right\}, 
\label{eq:1a}
\end{equation}
\begin{eqnarray}
\boldsymbol{S}_{i}^{(3)}\left(\boldsymbol{z}\right)&=&\left\{ S_{ijk}^{(3)}\left(\boldsymbol{z}\right),S_{ijl}^{(3)}\left(\boldsymbol{z}\right),S_{ijm}^{(3)}\left(\boldsymbol{z}\right),\cdots\right.\hphantom{********}\nonumber\\
&&\left.,S_{ikl}^{(3)}\left(\boldsymbol{z}\right),S_{ikm}^{(3)}\left(\boldsymbol{z}\right),\cdots,S_{ilm}^{(3)}\left(\boldsymbol{z}\right),\cdots\right\}. 
\label{eq:1b}
\end{eqnarray}
\end{subequations}
The elements of  $\boldsymbol{S}_{i}^{(3)}\left(\boldsymbol{z}\right)$ are independent syndromes, since $\boldsymbol{S}_{i}^{(3)}\left(\boldsymbol{z}\right)=\left\{ +1,+1,+1,\cdots\:,+1\right\} $ constitutes $\binom{N-1}{2}$ independent parity check equations.
Note that when $\boldsymbol{z}$ is code state, any physical spin variable $z_{jk}$ that is not involved in $\boldsymbol{z}_{i}$ can be derived from $N-1$ variables involved in $\boldsymbol{z}_{i}$ and
$\binom{N-1}{2}$ syndromes in $\boldsymbol{S}_{i}^{(3)}\left(\boldsymbol{z}\right)$ as $z_{jk}=z_{ij}z_{ik}S_{ijk}^{(3)}\left(\boldsymbol{z}\right)=z_{ij}z_{ik}$.
The set of the physical spin variables $\boldsymbol{z}_{i}$ is called logical lines associated with the logical spin $i$ \cite{Rocchetto2016,Fellner2022}. 
It corresponds to the available logical information, while the other spins give redundant information. 
Note also that $\boldsymbol{z}_{i}$ and its equivalent set $\tilde{\boldsymbol{z}}_{i}=\left\{ z_{ji},z_{ki},z_{li},z_{mi},\cdots\:\right\}$ correponds to $i$th row and column of the matrix $\boldsymbol{z}$. 

\begin{figure*}
\includegraphics[viewport=100bp 150bp 840bp 420bp,clip,scale=0.7]{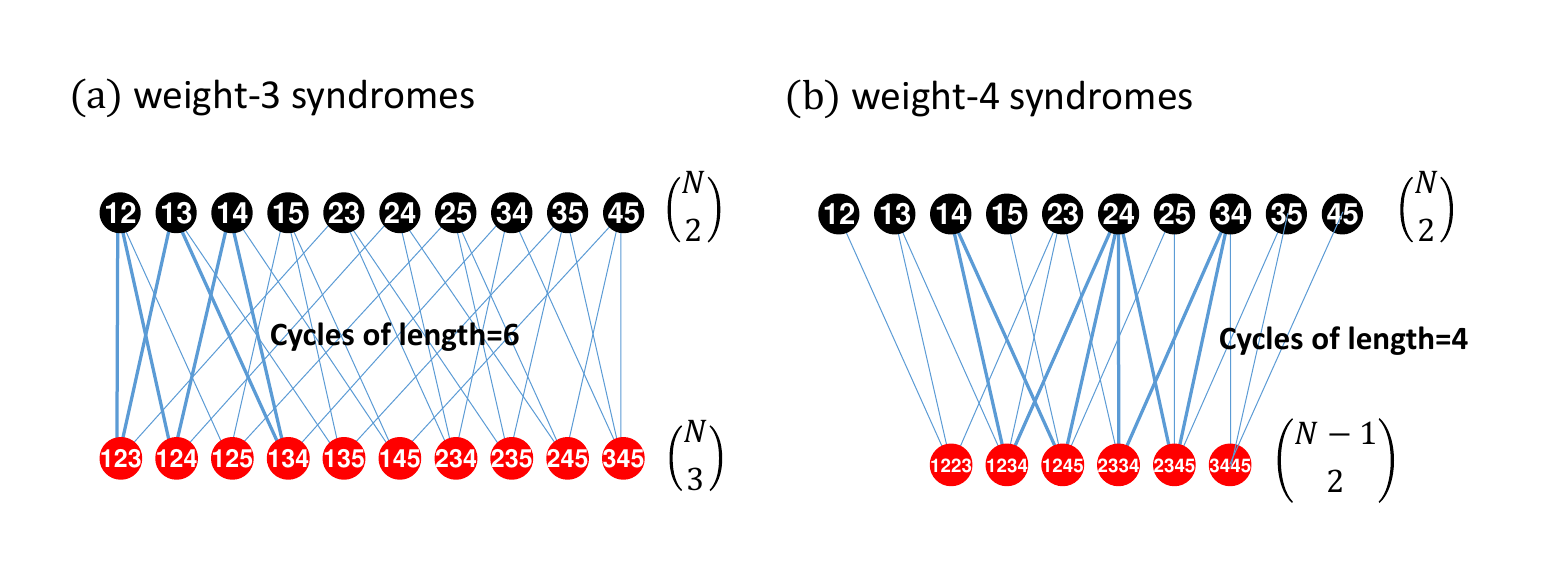}
\caption{
Bipartite graphs concerning the SLHZ system based on (a) weight-3 syndromes and (b) weight-4 syndromes.
The dark blue circle labeled with two integers $\left\{ i,j\right\} $ corresponds to the physical spin $z_{ij}$.
The red circle labeled with three integers $\left\{ i,j,k\right\} $ corresponds to weight-3 syndrome $S_{ijk}^{(3)}\left(\boldsymbol{z}\right)$, and the one labeled with four integers $\left\{ i,j,k,l\right\} $ corresponds to weight-4 syndrome $S_{ijkl}^{(4)}\left(\boldsymbol{z}\right)$. 
Following the graph model of the LDPC codes, the dark blue and red circles are called variable and check nodes, respectively.
\label{fig2}}
\end{figure*}
Figure \ref{fig2} shows the bipartite graphs of the SLHZ system based on (a) the weight-3 syndromes and (b) the weight-4 syndromes. 
Note that graph (b) is an irregular graph and is isomorphic to the graph shown in Fig.\ref{fig1}, i.e., $\tbinom{N}{2}$ physical spin variables (called variable nodes), indicated by dark blue circles, are placed on the upper side and $\tbinom{N-1}{2}$ syndromes  (check nodes), indicated by red circles, are placed on the lower side.
In contrast, Fig.\ref{fig2} (a) is a regular graph. 
In our algorithm, bipolar messages are iteratively exchanged between variable and check nodes until a solution is reached or the maximum number of iterations allowed is reached. 
As a result of the iterations, our algorithm becomes a kind of message-passing algorithm (MPA). 
This is crucial for decoding the readouts of the SLHZ system because, as PP pointed out, the SLHZ system is closely related to the LDPC codes \cite{Pastawski2016}. 
We point out that the length of shortest cycles of edges in Fig.\ref{fig2} (b) is 4, while that in Fig.\ref{fig2} (a) is 6. 
It is known that shorter cycles in the graph greatly reduce the convergence of the MPA. 
This is the reason why we chose the weight-3 syndrome and not the weight-4 syndrome for designing the decoding algorithm.

Let me explain our algorithm in more detail. 
Consider a general  $N\times N$ symmetric bipolar matrix $\boldsymbol{r}$ describing readouts of an annealing machine with possible errors. 
The matrix $\boldsymbol{r}$ must have unit diagonal elements, but off-diagonal elements $r_{ij}=r_{ji}\left(=\pm1\right)$ may not satisfy the parity constraints so that $S_{ijk}^{(3)}\left(\boldsymbol{r}\right)=-1$ for some sets $\left\{ i,j,k\right\}$. 
We define the inversion functions $\mathcal{F}_{ij}\left(\boldsymbol{r}\right)$ of the current readout $\boldsymbol{r}$ associated with physical spin $\left\{ i,j\right\}$ for any $1\leq i<j\leq N$:
\begin{equation}
\mathcal{F}_{ij}\left(\boldsymbol{r}\right)=1+\sum_{k\neq i,j}^{N} S_{ijk}^{(3)}\left(\boldsymbol{r}\right).
\label{eq:2}
\end{equation}
Note that $\mathcal{F}_{ij}\left(\boldsymbol{r}\right)$ is defined in terms of the weight-3 syndromes  $S_{ijk}^{(3)}\left(\boldsymbol{r}\right)$ $\left(1\leq k\neq i,j\leq N\right)$ adjacent to $r_{ij}$. 
Our algorithm's underlying assumption is that spin that is not correctly oriented will tend to have more unsatisfied syndromes adjacent to it than satisfied ones.
Thus,  $\mathcal{F}_{ij}\left(\boldsymbol{r}\right)$ can be regarded as a metric of the confidence in the correct orientation of $r_{ij}$. 
If $\mathcal{F}_{ij}\left(\boldsymbol{r}\right)<0$, enough of the set $A_{ij}=\left\{ S_{ij1}^{(3)}\left(\boldsymbol{r}\right),\cdots,S_{ijN}^{(3)}\left(\boldsymbol{r}\right)\right\} $ of $N-2$ syndromes adjacent to $r_{ij}$ are errorneous. 
Then, the sign of $r_{ij}$ is reversed to improve the confidence of the correct orientation. 
This algorithm is closely related to majority logic (MLG) decoding. 
The above rules can be rephrased as rules that update the current state $\boldsymbol{r}$ according to the rule 
\begin{eqnarray}
r_{ij}\rightarrow r_{ij}'
&=&r_{ij}\mathrm{sgn}\left(\mathcal{F}_{ij}\left(\boldsymbol{r}\right)\right)=\mathrm{sgn}\left(\mathcal{G}_{ij}\left(\boldsymbol{r}\right)\right)\nonumber\\
&=&\mathrm{maj}\left(r_{ij},r_{j1}r_{1i},\cdots,r_{jN}r_{Ni}\right),
\label{eq:3}
\end{eqnarray}
where 
\begin{equation}
\mathcal{G}_{ij}\left(\boldsymbol{r}\right)=r_{ij}\mathcal{F}_{ij}\left(\boldsymbol{r}\right)=r_{ij}+\sum_{k\neq i,j}^{N} r_{ik}r_{kj},
\label{eq:4}
\end{equation}
and $\mathrm{maj\left(\cdots\right)}$ denotes the majority function of bipolar variables \cite{Rudolph1972,Nambu2024}. 
The set $A_{ij}$ consists of $J=N-2$ syndromes orthogonal on error $e_{ij}=r_{ij}z_{ij}$ and the set $B_{ij}=\left\{ r_{ij},r_{j1}r_{1i},\cdots,r_{jN}r_{Ni}\right\}$ in the argument of the majority function consists of the associated $N-1$ estimators orthogonal on variable $z_{ij}$ \cite{Massey1962, Nambu2024}.
Eq.(\ref{eq:3}) is nothing but the one-step MLG decoding for $r_{ij}$ \cite{Nambu2024}. 
The orientations of all physical spins are estimated by simultaneously acting this map on the current $r_{ij}$ of all physical spins $\left\{ i,j\right\}$. 
It should be noted that the mapping of the variable $r_{ij}$ and the mapping of the variables $r_{ik}$ and $r_{kj}$ $(k\neq i,j)$ are not independent because $r_{ik}$ and $r_{kj}$ are included in the arguments of majority function in Eq. (\ref{eq:3}), causing the cycles of edges in the bipartite graph. 

\begin{figure}[tp]
\begin{centering}
\includegraphics[viewport=310bp 160bp 600bp 410bp,clip,scale=0.7]{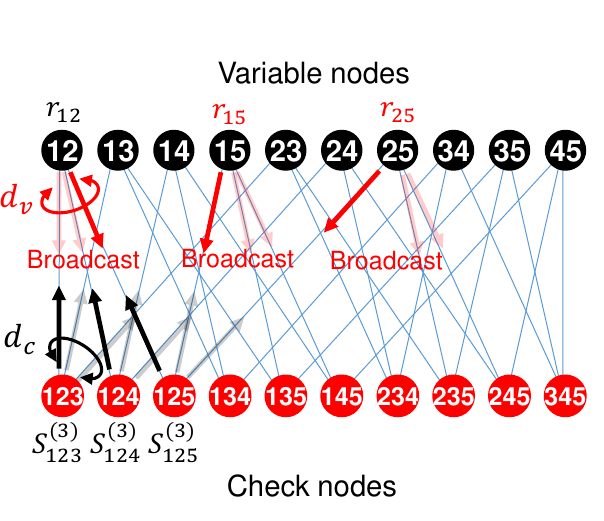}
\caption{
A bipartite graph to explain the principle of operation of our decoding algorithm.
\label{fig3}}
\end{centering}
\end{figure}
To decode the readout $\boldsymbol{r}$ of the SLHZ system, bipolar messages are exchanged between variable and check nodes. 
Let us assume that after the completion of the annealing calculation, the hard-decided, bipolar readout value $r_{ij}$ of each physical spin $\left\{ i,j\right\}$ is assigned as the initial value of the associated variable node. 
Under flipping, each variable node $\left\{ i,j\right\}$ broadcasts the $d_{v}=N-2$ adjacent check nodes $\left\{ i,j,k\right\}$ of its current value $r_{ij}$, and the check node $\left\{ i,j,k\right\}$ returns to the adjacent variable nodes $\left\{ i,j\right\}$ the parity $r_{ij}S_{ijk}^{(3)}\left(\boldsymbol{r}\right)=r_{ik}r_{kj}$ of the associated variable nodes excluding the returning variable node itself. 
Since $k\neq i,j$, there are $d_{v}=N-2$ such messages from any check nodes $\left\{ i,j,k\right\} $ to the associated variable node $\left\{ i,j\right\}$.
The current value $r_{ij}$ of each variable node $\left\{ i,j\right\}$ is then updated simultaneously by the majority of $r_{ij}$ and $r_{ik}r_{kj}$ $(k\neq i,j)$. 
See illustrative graph in Fig.\ref{fig3}, which focuses  the variable node $\left\{ i,j\right\} =\left\{ 1,2\right\}$ when $N=5$. 

Intuitively, if the update $\boldsymbol{r}'$ provides a better estimate of the target state $\boldsymbol{z}$ than the current $\boldsymbol{r}$, then the above process can be repeated to improve the quality of $\boldsymbol{r}$. 
This algorithm belongs to the standard MLG-based BF decoding of LDPC codes described in Ref.\cite{Zarrinkhat2004}, which is a sort of the Gallger's BF algorithm A \cite{Gallager1962, Gallager1963}. 
The important difference is that our algorithm is a multiple BF algorithm, while Gallager's is a single BF algorithm. 

\begin{figure}[tp]
\begin{centering}
\includegraphics[viewport=250bp 140bp 680bp 400bp,clip,scale=0.65]{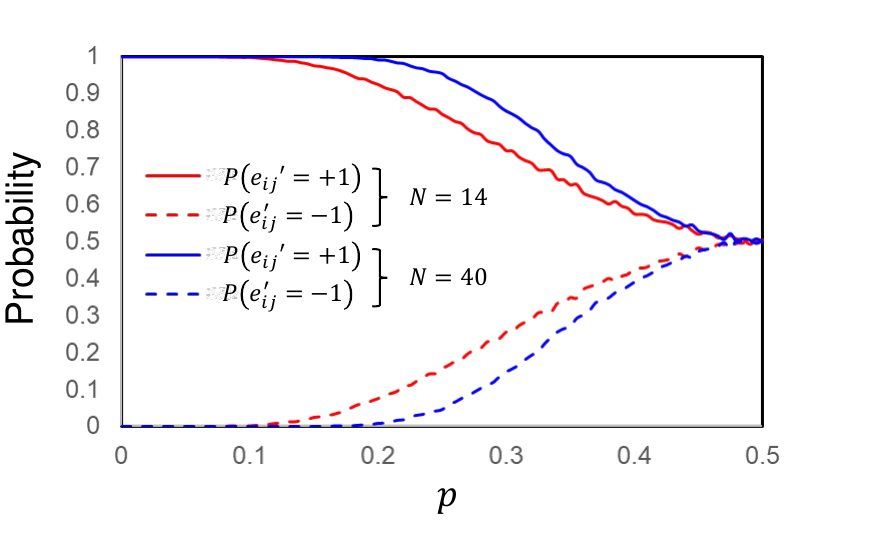}
\caption{
Probabilities $P\left(e_{ij}'=+1\right)$ ($P\left(e_{ij}'=-1\right)$)  for correctly (incorrectly) deciding variable $r_{ij}$ by decoding against error rate $p=P\left(e_{ij}=-1\right)$ for every spin orientation.
\label{fig:4}}
\end{centering}
\end{figure}
A comprehensive description of the principle of operation of this algorithm follows. 
We introduce an error pattern matrix. 
It is an $N\times N$ symmetric bipolar matrix, defined by $\boldsymbol{e}=\boldsymbol{r}\circ$$\boldsymbol{z}$, where $\boldsymbol{z}$ is a target code state satisfying $z_{ij}=Z_{i}Z_{j}$ for any $1\leq i,j\leq N$ and $\circ$ denotes element-wise (Hadamard) product of two matrices, i.e. $e_{ij}=r_{ij}z_{ij}\left(=\pm1\right)$. 
Since $\boldsymbol{z}$ is a code state, $S_{ijk}^{(3)}\left(\boldsymbol{z}\right)$ satisfies parity check equation and it follows that $S_{ijk}^{(3)}\left(\boldsymbol{r}\right)=S_{ijk}^{(3)}\left(\boldsymbol{e}\right)$.
Therefore, the syndrome $S_{ijk}^{(3)}\left(\boldsymbol{r}\right)$ is a function only of $\boldsymbol{e}$. 
It suffices to analyze errors of the SLHZ system for the ferromagnetically ordered target state $\boldsymbol{z}$ given by $z_{ij}=+1$ for any possible $\left\{ i,j\right\}$. 
This is because for arbitrary state $\boldsymbol{r}$ and the target state $\boldsymbol{z}$, the transformation $r_{ij}\rightarrow r_{ij}z_{ij}=e_{ij}$ results in the same energy of the SLHZ system if we also transform $J_{ij}\rightarrow J_{ij}z_{ij}$, where $J_{ij}$ is coupling constants for local fields in the SLHZ Hamiltonian as shown later.
Physically, this transformation is equivalent to redefining the axis of coordinates of the physical spins so that $\boldsymbol{z}$ is a ferromagnetically ordered state, which means a gauge invariance of the SLHZ system \cite{Rujan1993, Nishimori1993}.

It is easy to show that 
\begin{eqnarray}
\mathcal{F}_{ij}\left(\boldsymbol{r}\right)
&=&\mathcal{F}_{ij}\left(\boldsymbol{e}\right)=1+\sum_{k\neq i,j}^{N} S_{ijk}^{(3)}\left(\boldsymbol{e}\right)\nonumber\\
&=&e_{ij}\left(e_{ij}+\sum_{k\neq i,j}^{N} e_{ik}e_{kj}\right)=e_{ij}\mathcal{G}_{ij}\left(\boldsymbol{e}\right).
\label{eq:5}
\end{eqnarray}
Then, if $r_{ij}\rightarrow r_{ij}'$ is given by Eq. (\ref{eq:3}), it follows that 
\begin{equation}
e_{ij}\rightarrow e_{ij}'=\mathrm{sgn}\left(\mathcal{G}_{ij}\left(\boldsymbol{e}\right)\right)=\mathrm{maj}\left(e_{ij},e_{i1}e_{1j},\cdots,e_{iN}e_{Nj}\right).\label{eq:6}
\end{equation}
The probability $P\left(e_{ij}'=+1\right)$ of obtaining $e_{ij}'=+1$ after decoding was evaluated by applying Eq.(\ref{eq:6}) assuming a binary symmetric channel (BSC) model with independent and identical (i.i.d.) noise. 
Assume that the error probability of the readout $r_{ij}$ (called intrinsic information) is $P\left(e_{ij}=-1\right)=p$ for all physical spins $\left\{ i,j\right\}$. 
Using the lemma given by Massay and Gallager \cite{Massey1962,Gallager1962,Gallager1963}, the probability $P\left(e_{ik}e_{kj}=+1\right)=P_{e}$ that is common for all possible sets $\left\{ i,j,k\right\}$ (called extrinsic information) is given by $P_{e}=\frac{1}{2}\left[1+\left(1-2p\right)^{2}\right]$. 
Then, $P\left(e_{ij}'=+1\right)$ is given by the probability that the majority of the $N-1$ bipolar random variables are $+1$, where $+1$ occurs with probability $P\left(e_{ij}=+1\right)=1-p$ for variable $e_{ij}$ and occur with probability $P_{e}$ for the other $N-2$ variables $e_{ik}e_{kj}$ $(k\neq i,j)$. 
This probability can be evaluated by computer simulation of biased coin tossing. 
Fig.\ref{fig:4} shows the probability $P\left(e_{ij}'=+1\right)$ of correct deciding $r_{ij}$ and the probability $P\left(e_{ij}'=-1\right)$ of incorrect deciding it against $p$ evaluated after $10000$ trials of $N-1$ coin tosses.
In this figure, $P\left(e_{ij}'\right)$ for cases $N=14$ (red lines) and $N=40$ (blue lines) logical spins are shown as examples. 
The following results are obtained:
First, when $p<\frac{1}{2}$, the correct $r_{ij}$ can be decided with probability $P\left(e_{ij}'=+1\right)>\frac{1}{2}$. 
Second, $P\left(e_{ij}'=+1\right)$ is larger the larger $N$ is. 
These suggest that if $p<\frac{1}{2}$,  $r_{ij}$ for each physical spin can be correctly decided with probability greater than $\frac{1}{2}$ by the MLG-based mapping given by Eq.(\ref{eq:3}).
Therefore, this mapping may not necessarily decide the correct result for all entries of $\boldsymbol{r}$. 
Since the post-decoding probability $P\left(e_{ij}'=+1\right)$ increases as the pre-decoding probability $P\left(e_{ij}=+1\right)=1-p$ increases, iterative decoding is expected to improve the reliability of decision because fewer bits remain in error as the algorithm corrects errors at each iteration.
Therefore, by repeating the MLG mapping on $\boldsymbol{r}$ in Eq. (\ref{eq:3}), we can expect $\boldsymbol{r}$ to converge to target state $\boldsymbol{z}$.

\begin{figure*}
\begin{centering}
\includegraphics[viewport=78bp 100bp 800bp 450bp,clip,scale=0.66]{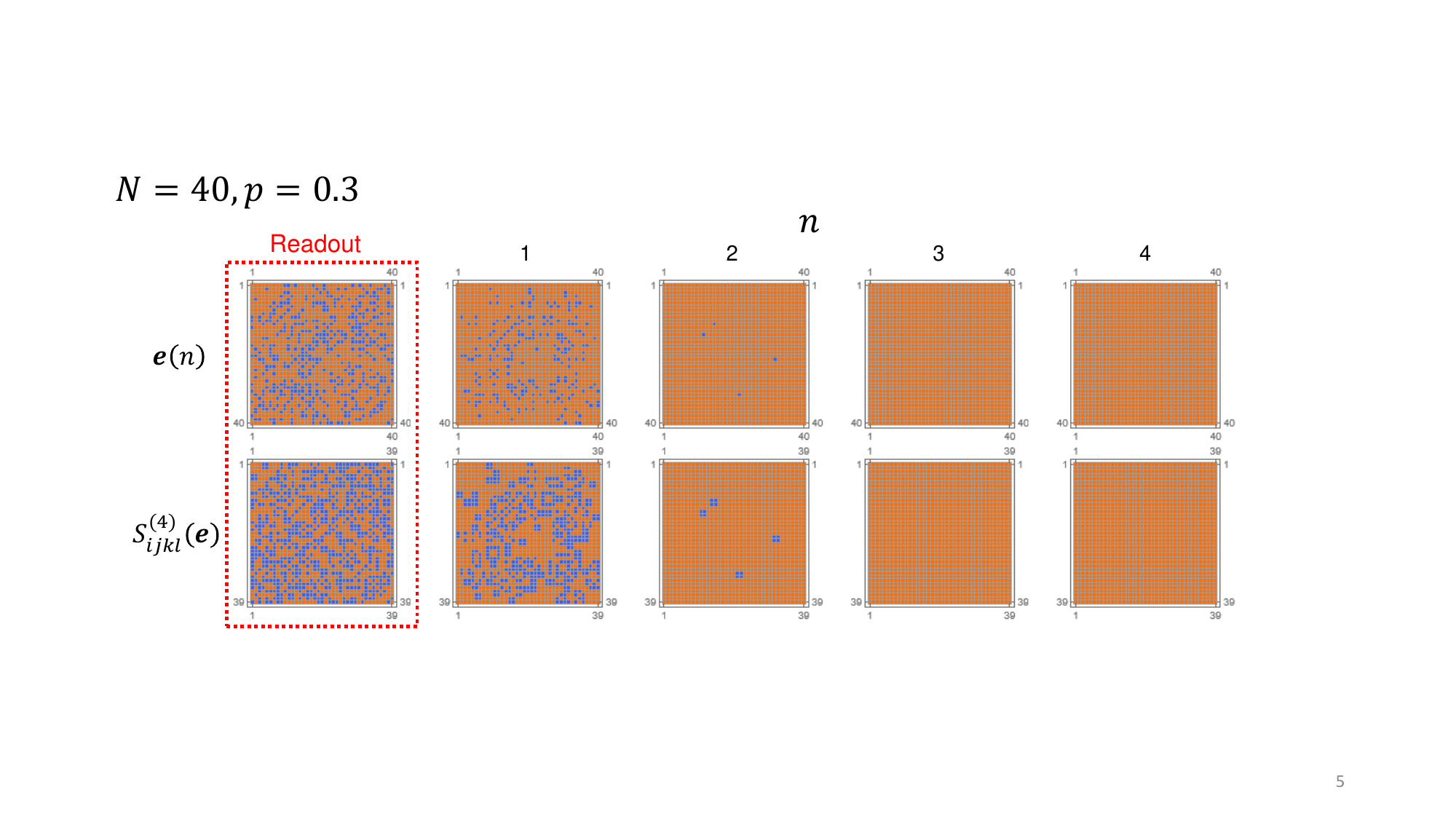}
\caption{
MLG-based BF decoding for $N=40$ and $p=0.3$. 
The upper figures show the readout $\boldsymbol{e}$ and decoded state after $n$ rounds of iteration of the map given by Eq. (\ref{eq:6}). 
The lower figures show the associated weight-4 syndrome patterns.
Blue pixels are spins or syndromes with errors. 
\label{fig:5}}
\end{centering}
\end{figure*}
\begin{figure*}
\begin{centering}
\includegraphics[viewport=100bp 180bp 780bp 440bp,clip,scale=0.75]{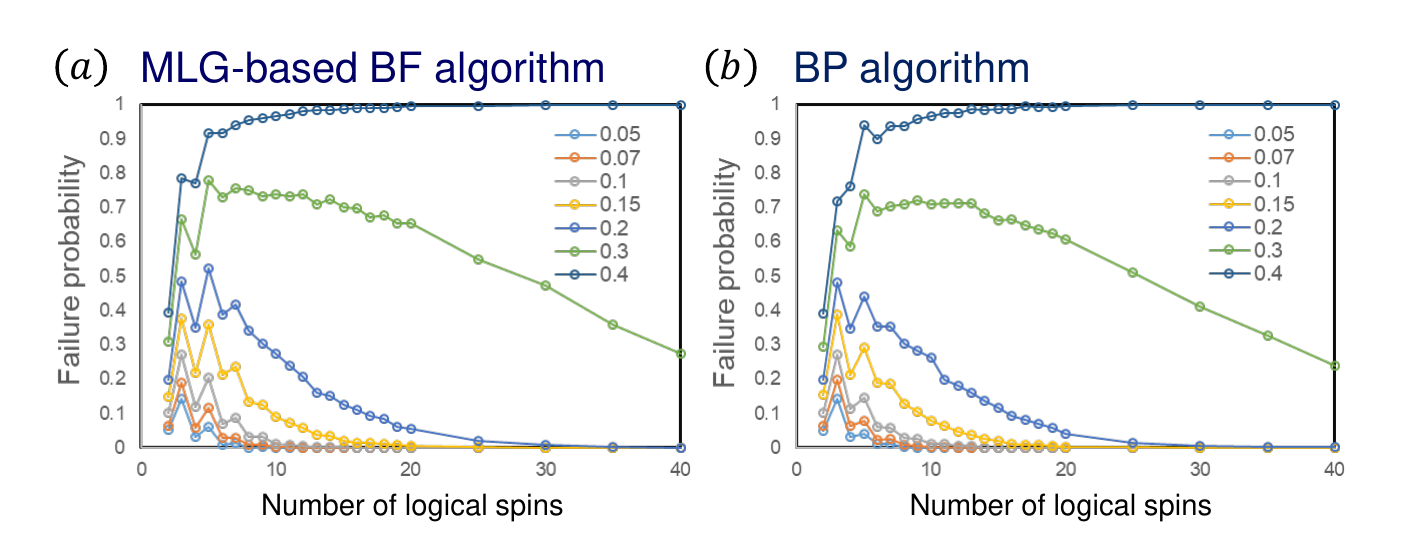}
\caption{
Dependence of decoding failure probability on the number of logical spins for (a) MLG-based BF algorithm and (b) BP algorithm for seven values of $p$ when assuming BSC and i.i.d. noise.
\label{fig:6}}
\end{centering}
\end{figure*}
The above intuition is actually confirmed by computer simulation. 
To prove the effectiveness of our algorithm in the SLHZ system, we generated an ensemble of $N\times N$ symmetric matrices $\boldsymbol{e}$ with unit diagonal elements and other elements assigned to $-1$ and $+1$ with the probability $p<\frac{1}{2}$ and $1-p>\frac{1}{2}$, respectively. 
This simulates a noisy readout of the SLHZ system assuming an i.i.d. noise model with error probability $p$ examined in the BP algorithm by PP \cite{Pastawski2016}.
The mapping from the current $\boldsymbol{e}\left(n\right)$ to next $\boldsymbol{e}\left(n+1\right)$ in Eq.(\ref{eq:6}) was applied to each entry of $\boldsymbol{e}$ simultaneously, where $n$ is the currrent round of iteration.  
Since the sum in the parentheses in Eq.(\ref{eq:6}) is equal to each entry of the matrix $\boldsymbol{e}\left(\boldsymbol{e}-\boldsymbol{I}\right)$, where $\boldsymbol{I}$ is an $N\times N$ identity matrix, this mapping can be done easily and efficiently using various matrix multiplication techniques. 
An example of the application of our decoding algorithm when $N=40$ and $p=0.3$ is shown in Fig.\ref{fig:5}. 
In this figure, each entry of the matrix $\boldsymbol{e}$ is plotted after $n=1,\cdots,4$ rounds of iteration of the mapping in Eq.(\ref{eq:6}). 
Blue pixels correspond to physical spins with errors, the number of which gradually decreases as rounds of iteration are added.
Finally, we could obtain an error-free matrix in $n\geq3$ rounds. 
It should be noted that the success probability for decoding is bounded and depends on $N$ and $p$. 
In this case, it was about $0.72$, which was confirmed as follows. 

\begin{figure*}
\begin{centering}
\includegraphics[viewport=130bp 130bp 800bp 420bp,clip,scale=0.75]{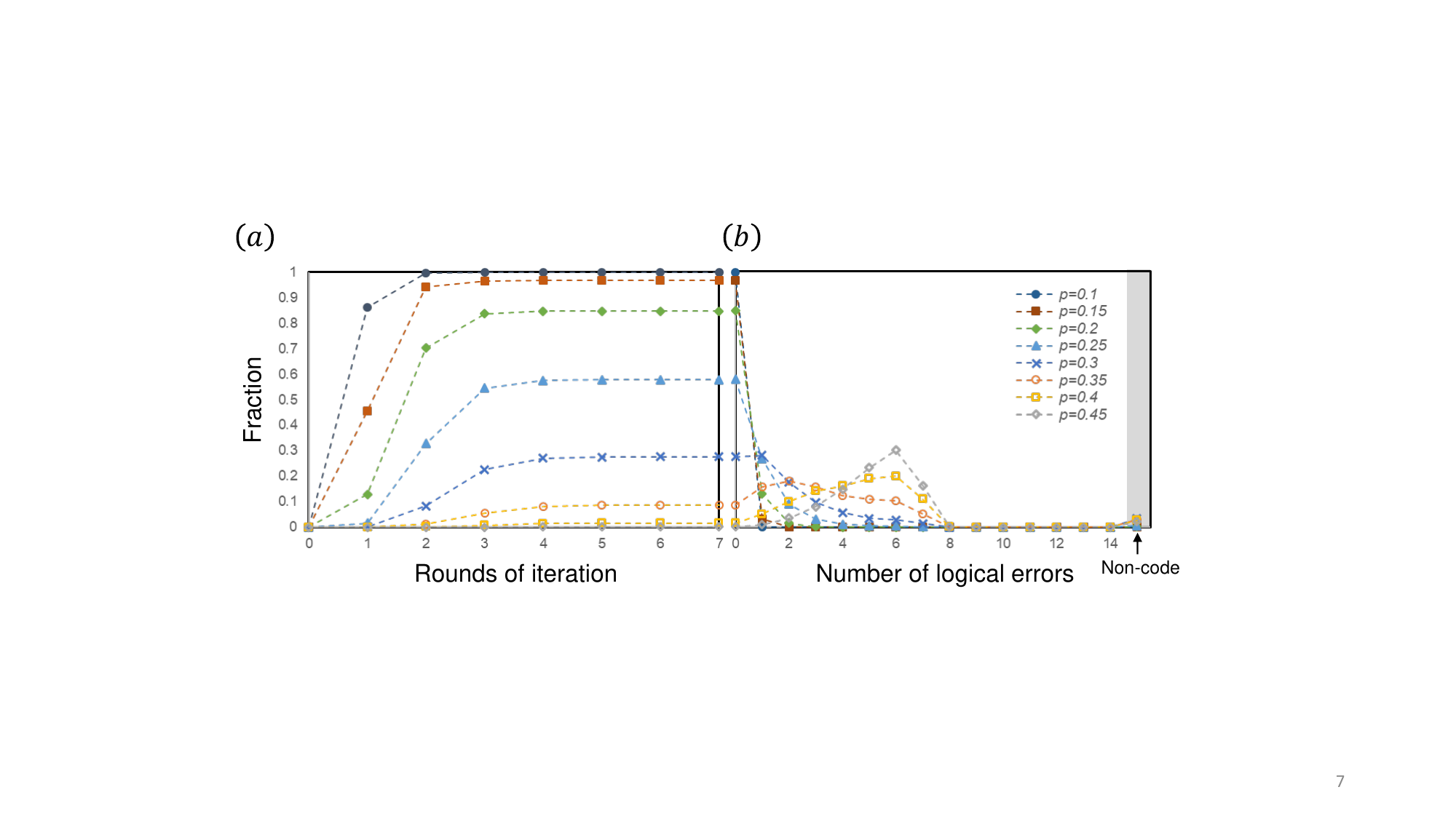}
\caption{
For eight values of $p$ $\left(=0.1,0.15,0.2,0.25,0.3,0.35,0.4,0.45\right)$, (a) a fraction of error-free states after decoding against the rounds of iteration and (b) a fraction of code states after decoding of seven rounds of iteration against the number of logical errors in those states, respectively. 
The data indicated by ``Non-code'' in the right figure shows a fraction of states that were not decoded to code states after decoding.
\label{fig:7}}
\end{centering}
\end{figure*}

The performance of this algorithm was evaluated by randomly generating a large number of initial error matrices $\boldsymbol{e}$ and checking whether the errors are eliminated by decoding. 
The number of error pattern matrices $\boldsymbol{e}$ was $5000$, and the number of iterations was $5$. 
Fig.\ref{fig:6} (a) shows the failure probability of decoding which is plotted for $N$ ranging from $2$ to $40$ for seven values of
$p$ $\left(=0.05,0.07,0.1,0.15,0.2,0.3,0.4\right)$. 
This result should be contrasted with the performance of BP algorithm in Fig.\ref{fig:6} (b) given by Ref.\cite{Pastawski2016}, which is essentially a sum-product algorithm in the probability domain. 
It is evident that as long as assuming BSC and i.i.d. noise, the performance of our decoding algorithm is comparable to that of BP algorithm. 
Fig.\ref{fig:7} shows the further features of our decoding algorithm. 
The figure shows, for eight values of  $p$ $\left(=0.1,0.15,0.2,0.25,0.3,0.35,0.4,0.45\right)$, (a) a fraction of error-free state against the number of rounds of iteration and (b) a fraction of logical state (with and without logical errors) obtained after seven rounds of iteration, against their associated number of logical errors. 
Fig.\ref{fig:7} (a) shows that as the round of iteration increases, a fraction of the error-free state increases and saturates at five rounds. 
We also see that the saturated value decreases with increasing $p$. 
In contrast, Fig.\ref{fig:7} (b) shows that the decoded states are most likely code states, but the number of logical errors increases as $p$ increases. 

It should be noted that our decoding algorithm is much simpler than the BP algorithm. 
This is primarily because, in effect, each edge coming out of a variable node carries the same broadcast message. 
The second reason is that the message exchanged between adjacent nodes at each iteration has only bipolar values. 
Furthermore, our algorithm has peculiar feature that the functions $\mathcal{G}_{ij}\left(\boldsymbol{r}\right)$ in Eq.(\ref{eq:3}) can be readily calculated by matrix calculation $\boldsymbol{r}\left(\boldsymbol{r}-\boldsymbol{I}\right)$. 
In fact, using recent tensor computing techniques would calculate these with very large throughput.

\section{Error-correcting capability for thermal samples\label{sec:3}}

So far, we have focused on the performance of our decoding algorithm, assuming an i.i.d. noise model in SLHZ system readouts.
Even if this model is adequate for errors that occur when the ground state is observed using independent imperfect measurement apparatuses, it may be inadequate for quantum/classical annealing. 
This is because errors may arise from states ultimately excited by many spin flips that differ from the ground state. The spin flips may be neither random nor weakly correlated in the closed or open system.
Albash et al.\cite{Albash2016} provided evidence that this noise model is inadequate by investigating the performance of the SLHZ scheme using SQA, a Monte Carlo method that approximates the behavior of the QA. 
In addition to this inadequacy, the discussion in the previous section implicitly assumed the existence of a priori knowledge of noise parameters such as error probabilities. 
Such parameters can be evaluated in advance for communication applications but are impossible to obtain for annealing calculations. 
Therefore, it is important to study how and to what extent our decoding algorithm, when used in conjunction with annealing, contributes to solving optimization problems.

Since QA device based on the SLHZ system is still in the developmental stage, it is difficult to test our decoding algorithm on an actual QA device. 
Instead, they were tested using classical Markov chain Monte Carlo (MCMC) simulations. 
This would be reasonable because several studies suggest that the currently available QA may function well as a fast thermal sampler \cite{Marshall2017,Sarandy2005, Sarovar2013, Amin2015, Benedetti2016}. 
QA is expected to potentially sample from an MB-like distribution of Hamiltonians more efficiently than classical methods \cite{Benedetti2016}.
The temperature of the distribution is considered to be that of the freeze-out point in the dynamical evolution of annealing \cite{Amin2015,Benedetti2016,Marshall2019}. 
Since the canonical MCMC algorithm can sample from MB distribution, our analysis may help to understand the validity of our decoding algorithm and how it will be used for post-processing for QA based on the SLHZ system. 

The ensemble of the readouts was prepared as follows.
The Hamiltonian of the SLHZ system is written as \cite{Nambu2024}
\begin{equation}
H^{SLHZ}\left(\boldsymbol{z}\right)=-\beta\sum_{\left\{ i,j\right\}}J_{ij}z_{ij}+\gamma\sum_{plaquettes}\frac{1-S_{ijkl}^{(4)}\left(\boldsymbol{z}\right)}{2},
\label{eq:7}
\end{equation}
where the sum in the first term is over all physical spins $\left\{ i,j\right\} $, the sum in the second term is over all plaquettes, i.e., the unit cell of physical spins for which the weight-4 syndrome is defined, and $\beta,\gamma>0$ are real weight parameters that adjust the relative emphasis between the two terms. 
The first term corresponds to the correlation between the set of bipolar variables $\left\{ z_{ij}\right\} $ and the given set of real coupling parameters $\left\{ J_{ij}\right\} $. 
Denoting $\left\{ z_{ij}\right\}$ as an $N\times N$ symmetric bipolar matrix $\boldsymbol{z}$, we can denote $\left\{ J_{ij}\right\} $ as an $N\times N$ symmetric real matrix $\boldsymbol{J}$, and the first sum can be written as the matrix inner product $\left\langle \boldsymbol{J},\boldsymbol{z}\right\rangle$.
The second term is proportional to the sum of the weight-4 syndromes.
Let the vector $\boldsymbol{Z}=\left(Z_{1},\cdots,Z_{N}\right)^{T}$ represent a state of the $N$ logical spins and let $\tilde{C}$ represent the space of code state $\boldsymbol{z}=\boldsymbol{Z}\otimes\boldsymbol{Z}$ (called code space).
Then, if and only if $\boldsymbol{z}\in\tilde{C}$, the second term has its minimum value $0$. 
Thus, the second term can be considered as a penalty term, which forces $\boldsymbol{z}$ to be a valid code state. 
Note that this Hamiltonian is a non-linear function and has many local minima. 
These local minima are a major obstacle to finding the solution below.

The solution of  
\begin{equation}
\tilde{\boldsymbol{z}}=\underset{\boldsymbol{z}\in\tilde{C}}{\arg\min}H^{SLHZ}\left(\boldsymbol{z}\right)
\label{eq:8}
\end{equation}
gives us the solution of the global optimization of the logical Hamiltonian
\begin{equation}
H^{log}\left(\boldsymbol{Z}\right)=-\sum_{\left\{ i,j\right\}}J_{ij}Z_{i}Z_{j},
\label{eq:9}
\end{equation}
that is 
\begin{equation}
\tilde{\boldsymbol{Z}}=\underset{\boldsymbol{Z}}{\arg\min}H^{log}\left(\boldsymbol{Z}\right),
\label{eq:10}
\end{equation}
where $\tilde{\boldsymbol{z}}=\tilde{\boldsymbol{Z}}\otimes\tilde{\boldsymbol{Z}}$.
In the communication application scenario, $\left\{ J_{ij}\right\} $ is the soft-decision that is sampled at the receiver output. 
In this scenario, finding the solution $\tilde{\boldsymbol{z}}$ corresponds to maximum a posteriori (MAP) decoding of an LDPC code using weight-4 syndromes \cite{Nambu2024}.
Typically, $\left\{ z_{ij}\right\} $ is initialized as a hard decision of $\left\{ J_{ij}\right\} $, i.e. $z_{ij}^{(0)}=\mathrm{sgn\left(\mathit{J_{ij}}\right)}$ for possible set of $\left\{ i,j\right\} $. 
Thus, MAP decoding is to find the decision $\boldsymbol{z}\in\tilde{C}$ that has a maximum correlation with the received samples $\left\{ J_{ij}\right\}$.
The allowed solution $\boldsymbol{z}$ must be in code state. 
Under this constraint, the solution that minimizes the Hamiltonian (\ref{eq:7}) is also a solution to MAP decoding defined by (\ref{eq:10}).
In contrast, BP and our MLG-based BF algorithm are based on the maximizer of posterior marginals (MPM) decoding, which is a symbol-wise version of MAP decoding \cite{Nambu2024}.

To find the globally optimum solution $\tilde{\boldsymbol{z}}$, we have been using classical RFMCMC simulation \cite{Nambu2022,Nambu2024}. 
RFMCMC is an MCMC method for obtaining random sample sequences from a priori probability distributions that are difficult to sample directly. 
It is known that canonical MCMC uising the Metropolis acceptance-rejection rule can sample from the MB distribution. 
Random sampling from an ensemble associated with a low-temperature distribution with a high probability of occupancy of the ground state would be expected to sample the ground state with a high probability.
However, the stationary distribution is highly dependent on self-loop transitions derived from the diagonal elements of the transition kernel of the canonical MCMC. 
To lower the equilibrium temperature of the stationary distribution, the magnitude of the diagonal element of the kernel associated with the low-lying energy state must be increased.
Unfortunately, self-loop transitions associated with low-lying excited states result in slow mixing and make ground-state sampling inefficient because the Markov chain is stuck in such low-lying excited states most of the time.
This is quite reasonable if one notes that self-loop transitions correspond to rejection events associated with the Metropolis rule, which is conventionally interpreted as trapping of the spins to local energy minima. 
Therefore, efficient sampling is a difficult task for canonical MCMC. 
In contrast, the RFMCMC has no diagonal elements in its transition kernel and no self-loop transitions. 
This results in inefficient sampling, as the probability of occupancy of the ground state in the stationary distribution becomes small. 
Nevertheless, bookkeeping algorithms that track the best solution help resolve this problem since they can avoid time-wasting self-loop transitions without degrading performance \cite{Nambu2022,Nambu2024}.
It should also be noted that the stationary distribution of the RFMCMC is the MB distribution, excluding the state multiplicity due to self-loop transitions \cite{Rosenthal2021, Nambu2024}.
The samples in the canonical MCMC reflect multiplicity due to self-loop transition, $M_{k}\sim\mathrm{\mathrm{1+Geometric}}\left(\alpha_{0}\left(k\right)\right)\geq1$, for every $k$th state that is a function of acceptance rate $\alpha_{0}\left(k\right)$ for the $k$th state.
In contrast, $M_{k}=1$ in the latter \cite{Nambu2024}. 
Ultimately, RFMCMC is a very efficient way to sample states that could potentially be sampled by canonical MCMC, that is, a thermal sampler. 
Since real-world QA is considered a fast thermal sampler \cite{Marshall2017,Sarandy2005,Sarovar2013,Amin2015,Benedetti2016}, this implies that if our decoding algorithm is valid for RFMCMC simulation, it is also valid for actual QA devices. 

\begin{figure*}
\begin{centering}
\includegraphics[viewport=60bp 100bp 900bp 490bp,clip,scale=0.62]{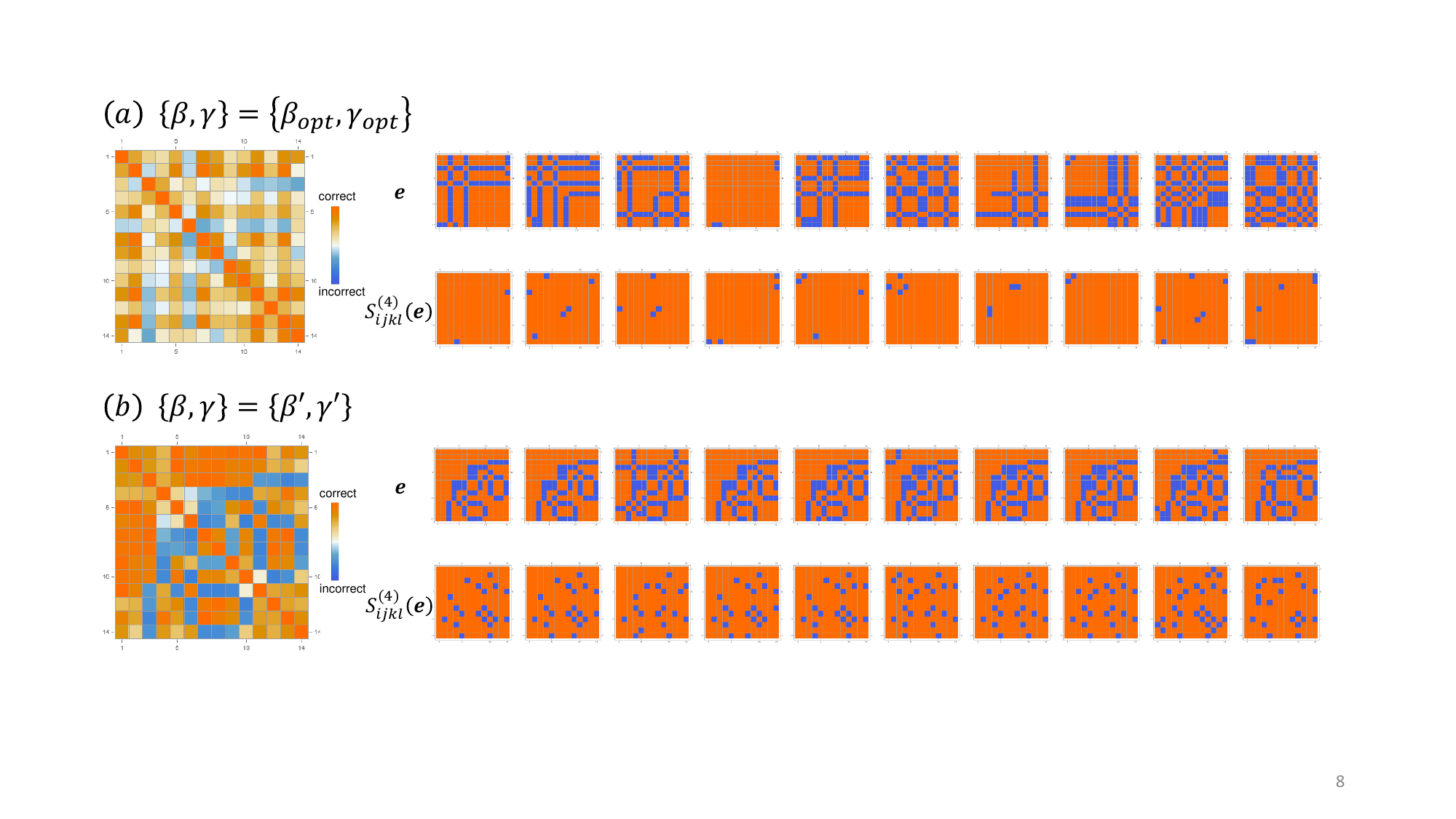}
\caption{
(Left) error probability distribution of the ensemble $\left\{ \boldsymbol{e}\right\}$ sampled by RFMCMC for cases (a) $\left\{ \beta,\gamma\right\} =\left\{ \beta_{opt},\gamma_{opt}\right\}$ and (b) $\left\{ \beta,\gamma\right\} =\left\{ \beta',\gamma'\right\}$, where the orange (blue) pixel indicates the correct (incorrect) spin orientation and its shading indicates the magnitude of the probability. 
(Right) ten typical samples of $\left\{ \boldsymbol{e}\right\}$ and associated weight-4 syndrome patterns.
\label{fig:8}}
\end{centering}
\end{figure*}
\begin{figure*}
\begin{centering}
\includegraphics[viewport=130bp 20bp 810bp 520bp,clip,scale=0.7]{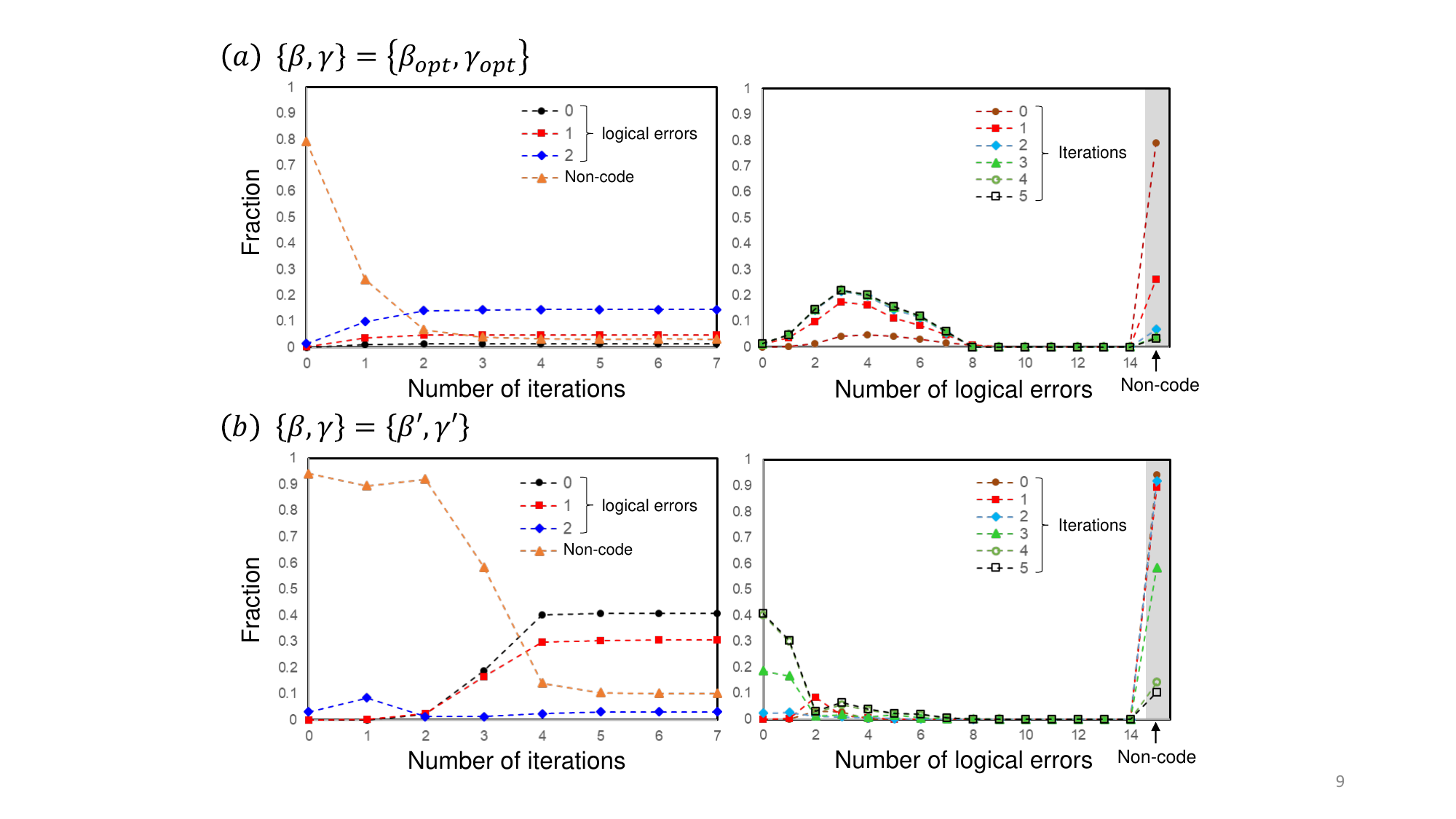}
\caption{
Performance of our decoding algorithm evaluated for cases of (a) $\left\{ \beta,\gamma\right\} =\left\{ \beta_{opt},\gamma_{opt}\right\}$ and (b) $\left\{ \beta,\gamma\right\} =\left\{ \beta',\gamma'\right\}$. 
The left traces show a fraction of states decoded to code states with logical errors 0, 1, and 2 and a fraction of states not decoded to code states after from $0$ (readout state) to $7$ rounds of iteration. 
The right trace shows a fraction of states decoded to code states with various logical errors and a fraction of states not decoded to code states after from $0$ to $5$ rounds of iteration. 
The data designated by "Non-code" in the right figures show a fraction of the states that were not decoded to the code states after decoding. 
\label{fig:9}}
\end{centering}
\end{figure*}
\begin{figure*}
\begin{centering}
\includegraphics[viewport=140bp 80bp 780bp 470bp,clip,scale=0.75]{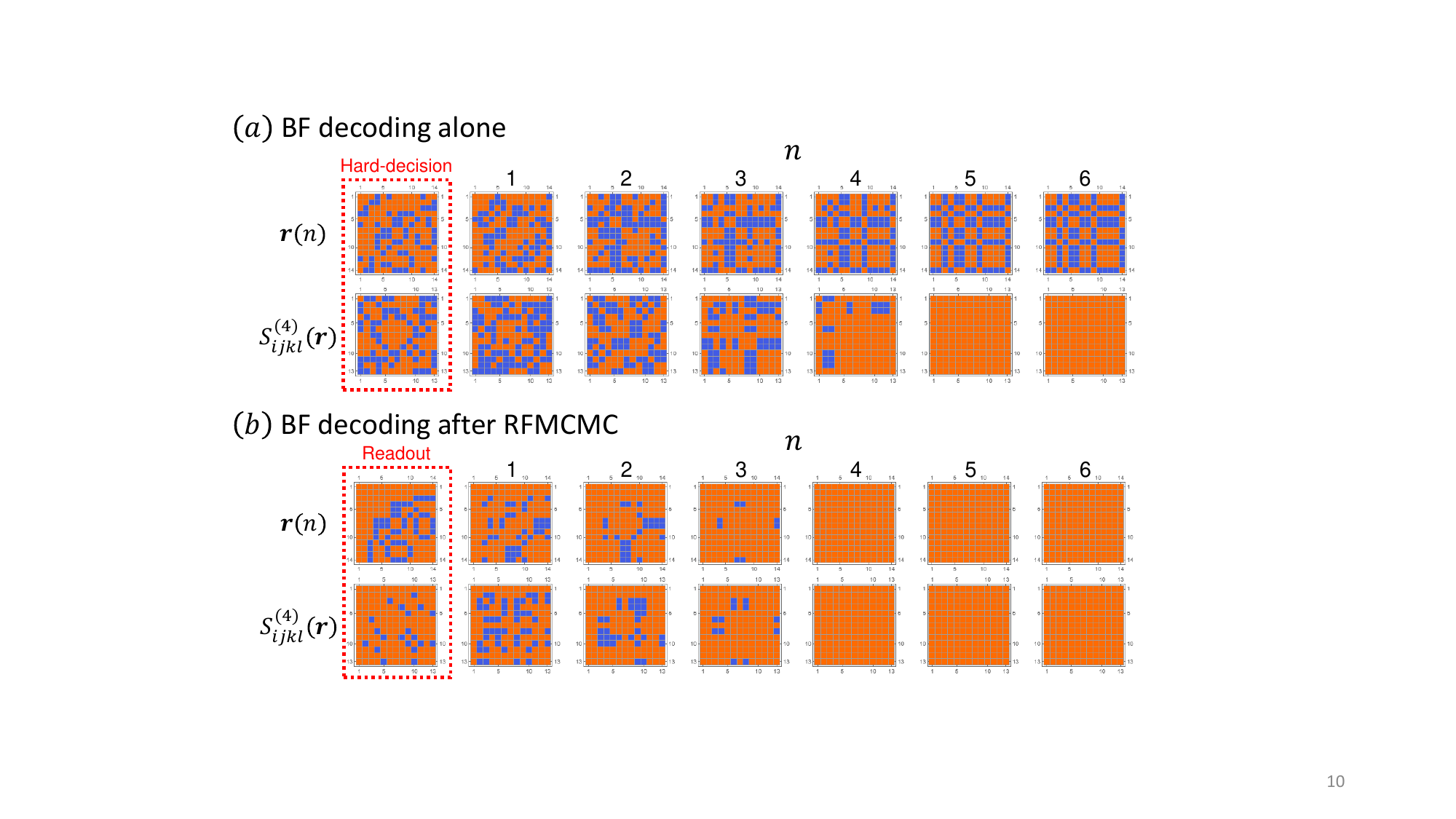}
\caption{
Decoded results after $n$ rounds of iteration, assuming the following initial conditions: (a) hard-decision given by Eq. (\ref{eq:11}) and (b) typical readout of RFMCMC.
The leftmost matrix plots indicate the assumed initial states.
\label{fig:10}}
\end{centering}
\end{figure*}
We implemented RFMCMC and investigated the performance of our decoding algorithm on readout samples created by the thermal process under two conditions. 
The success probability $p_{suc}$ of finding a unique globally optimal solution $\tilde{\boldsymbol{z}}$ for the spin glass problem of $N=14$ logical spin system was evaluated. 
The associated SLHZ system consists of $\tbinom{N}{2}=91$ physical spins. 
As an example, we chose a spin-glass problem in which each entry in the coupling matrix $\boldsymbol{J}$ was chosen from a uniformly distributed random variable in the range $[-1/4,1/4]$ and the solution $\tilde{\boldsymbol{z}}$ is known by a brute force calculation. 
For details of the simulation, see Ref.\cite{Nambu2022,Nambu2024}. 
Beforehand, an ensemble of many readouts  $\left\{ \boldsymbol{r}\right\}$ was prepared for various sets of static weight parameters  $\left\{ \beta,\gamma\right\} $ by a chain of RFMCMC for the SLHZ system.
For all parameter sets $\left\{ \beta,\gamma\right\} $, the cardinality of $\left\{ \boldsymbol{r}\right\} $ was 109200. 
We evaluated $p_{suc}$ as a function of a set of the parameter sets, $\left\{ \beta,\gamma\right\} $, and found that optimal parameter set $\left\{ \beta_{opt},\gamma_{opt}\right\} $ that maximizes $p_{suc}$ when the prepared readout $\boldsymbol{r}$ was observed without post-readout decoding. 
We also chose another parameter set $\left\{ \beta',\gamma'\right\} $ with $\beta'>\beta_{opt}$ and $\gamma'<\gamma_{opt}$, and prepared associated readout  ensemble $\left\{\boldsymbol{r}\right\}$. 
Finally, the known solution $\tilde{\boldsymbol{z}}$ was used to convert readout ensemble $\left\{\boldsymbol{r}\right\}$ to error ensemble $\left\{\boldsymbol{e}\right\}$. 
Figure \ref{fig:8} shows the statistical properties of the ensemble $\left\{ \boldsymbol{e}\right\}$ for cases (a) $\left\{ \beta,\gamma\right\} =\left\{ \beta_{opt},\gamma_{opt}\right\} $ and (b) $\left\{ \beta,\gamma\right\} =\left\{ \beta' \gamma'\right\}$.
The left and right plots are matrix plots of the error probability distribution of the ensemble $\left\{ \boldsymbol{e}\right\}$ and ten samples randomly chosen from $\left\{ \boldsymbol{e}\right\}$, respectively.
The error and syndrome patterns are found to be highly dependent on the weight parameters $\left\{ \beta,\gamma\right\}$. 
In the case (a), the success probability of finding error-free readouts, i.e., finding a globally optimal solution, was about $4.67\times10^{-4}$, 
while the success probability of finding error-free syndromes, i.e., finding code states, was about $2.29\times10^{-2}$. 
In contrast, in the case (b), both probabilities are zero; that is, readouts $\boldsymbol{r}$ all remained in non-code states. 
Comparing the right plots, the error patterns are irregular in the case (a), while they are relatively regular in the case (b). 
In addition, there are fewer syndrome errors in the case (a) than in the case (b).
These differences stem from differences in the thermal energy spectrum of the Hamiltonian. 

To test our decoding algorithm, we applied it to two randomly selected samples from each of the two readout ensembles described above. 
The size of the selected samples was $10000$. 
Figure \ref{fig:9} shows the performance of our decoding algorithm for the cases (a) and (b). 
The left traces show a fraction of 0, 1, and 2 logical error states and non-code states obtained after decoding from $0$ (readout state) to $7$ rounds of iteration. 
The right trace, on the other hand, shows a fraction of various logical error states and non-code states obtained after decoding from $0$ to $5$ rounds of iteration. 
In the case (b), we see that a large fraction of the prepared readouts are decoded to an error-free state. 
Conversely, in the case (a), most of the readouts are decoded to code states but with logical errors, which our decoding algorithm cannot correct.
Thus, the performance of our decoding algorithm depends on the weight parameter set $\left\{ \beta,\gamma\right\} $, and surprisingly, it is better for case $\left\{ \beta,\gamma\right\} \neq\left\{ \beta_{opt},\gamma_{opt}\right\}$ than case $\left\{ \beta,\gamma\right\} =\left\{ \beta_{opt},\gamma_{opt}\right\}$. 
The low performance in the case $\left\{ \beta,\gamma\right\} =\left\{ \beta_{opt},\gamma_{opt}\right\}$ may be due to the large fluctuation in error patterns as shown in Fig.\ref{fig:8}. 
In fact, the error probability per spin averaged over all the physical spins was about 0.39 in the case (a), while it was about 0.31 in the case (b). 
This result suggests that large fluctuations in the annealing process are necessary to find the globally optimal solution $\tilde{\boldsymbol{z}}$, but unfortunately, they degrade the performance of the decoding algorithm.
On the other hand, in the case (b), the error probability is somewhat smaller, but the performance of the decoding algorithm is still significant.
Note that the error patterns in Fig.\ref{fig:8} are not random as expected in an i.i.d. noise model.
These patterns reflect the thermal energy spectrum of the Hamiltonian (\ref{eq:7}), which depends on the choice of weight parameters $\left\{ \beta,\gamma\right\}$, but do not reflect the dynamical evolution of the SLHZ system \cite{Amin2015}. 

It is worth noting that the decoding contribution is large in the case (b), which is far from the optimal condition $\left\{ \beta,\gamma\right\} =\left\{ \beta_{opt},\gamma_{opt}\right\}$ if an RFMCMC simulation alone is used to find a globally optimal solution $\tilde{\boldsymbol{z}}$. 
This is quite reasonable if we consider the fact that the success probability for our decoding algorithm depends on its assumed initial state, 
that is, decoding succeeds as long as the initial state is correctable. 
For example, consider the scenario of a communication application.
In this scenario, finding a globally optimal solution given by Eq.(\ref{eq:8}) amounts to the MAP decoding, 
where we usually assume the initial state $\boldsymbol{r}^{(0)}$ as the hard-decision of soft-decided variables $\left\{ J_{ij}\right\}$, i.e.,
\begin{equation}
r_{ij}^{(0)}:=\mathrm{sgn}\left(J_{ij}\right)
\label{eq:11}
\end{equation}
for every physical spin $\left\{ i,j\right\}$, which minimizes only the correlation term but not the penalty term in  $H^{SLHZ}\left(\boldsymbol{z}\right)$ \cite{Wadayama2007,Yatribi2020}. 
Then, finding the globally optimal solution is equivalent to finding the closest code state to $\boldsymbol{J}$. 
Figure \ref{fig:10} (a) shows the results of our decoding algorithm after $n$ rounds of iteration for the initial state given by  Eq.(\ref{eq:11}).
The leftmost matrix plot shows the initial state $\boldsymbol{r}^{(0)}$ assumed in Eq.(\ref{eq:11}). 
After $n=5$, the decoded state converges to the code state corresponding to the five logical error states.
This indicates that the assumed initial state is not correctable by our decoding algorithm. 
It should be noted that this is not always the case.
For example, if we consider the optimization problem of inverting $\boldsymbol{J}$, i.e., $\boldsymbol{J}\rightarrow\boldsymbol{-J}$, we confirm that the associated initial state $\boldsymbol{r}^{(0)}$ successfully converges to its error-free state by our decoding algorithm. 
Thus, when applied to the hard-decided initial state $\boldsymbol{r}^{(0)}$ given by Eq. (\ref{eq:11}), the performance of our decoding algorithm depends on $\boldsymbol{J}$. 

The above results suggest that the performance of our decoding algorithm is limited when used as a stand-alone solver for the globally optimal solution. 
We can consider two shortcomings leading to this limitation. 
First, we did not use soft information on the reliability of the hard-decided variables $\left\{ r_{ij}^{(0)}\right\}$ which should be dependent on the soft-decided parameter sets $\left\{ J_{ij}\right\}$. 
This means that our algorithm implicitly assumed uniform reliability without considering noise properties.
Second, because our algorithm is deterministic, it tends to get stuck at local energy minima, preventing the solution from converging to an error-free globally optimal solution. 
These problems have been heuristically avoided by incorporating weight parameters that depend on $\left\{ J_{ij}\right\}$ to take account of inhomogeneity in the reliability of $\left\{ r_{ij}^{(0)}\right\}$ \cite{Kou2001} as well as probabilistic spin flips to escape from undesirable local optima \cite{Miladinovic2005, Sundararajan2014} in the BF algorithms. 
We point out here that our decoding algorithm, supplemented by annealing calculations, provides an alternative to these sophisticated BF decoding algorithms. 
So far, decoding has been regarded as a post-process of annealing calculations. 
Conversely, we can consider that the performance of decoding can be improved by incorporating the annealing calculation as a pre-process of the decoding algorithm. 
In fact, annealing calculations provide a way to compensate for the above shortcomings. 
Annealing calculations can reflect the inhomogeneity of the magnitude of $\left\{ J_{ij}\right\}$ through the correlation term in $H^{SLHZ}\left(\boldsymbol{z}\right)$ and can probabilistically escape from local energy minima.
An annealing calculation allows the spin state to evolve from an uncorrectable initial state to a correctable readout if the weight parameters $\left\{ \beta,\gamma\right\}$ are properly adjusted, resulting in the globally optimal solution shown in Fig.\ref{fig:10} (b) after decoding.
Our classical BF decoding strategy after the annealing calculation is a hybrid calculation that sequentially uses probabilistic and deterministic algorithms.
To achieve high performance, we need to find the optimal setting for parameters $\left\{ \beta,\gamma\right\}$ that does not necessarily coincide with $\left\{ \beta_{opt},\gamma_{opt}\right\}$. 
It is worth noting that such hyperparameter optimization is also necessary for weighted BF algorithm \cite{Kou2001} and probabilistic BF algorithm \cite{Miladinovic2005} or noise-assisted BF algorithm  \cite{Sundararajan2014}.

\section{Conclusions}

We revisited the error correction problem for encoded QA based on the parity-encoding (SLHZ) system, which embodies a classical LDPC code. 
We proposed a simple deterministic decoding algorithm based on the BF decoding algorithm using weight-3 syndromes. 
Our algorithm's performance was comparable to that of the BP algorithm discussed by PP, assuming an i.i.d. noise model, even though the decoding complexity is much lower.
However, it is not obvious that our algorithm can correct errors that occur in QA. 
Since the error pattern is determined by the final-time state and the current QA results in an MB-like distribution of readouts, the error due to the thermal excitation process was investigated.
RFMCMC simulations were used to sample possible error patterns in the thermal excitation process, as they have been shown to converge to a stationary distribution derived from the MB distribution.
Two sets of weight parameters were chosen for the simulation, and the ensembles of readouts associated with them were sampled. 
Applying our decoding algorithm to these samples, we found that it is valid for parameter sets for which RFMCMC alone can not sample error-free state or even code state at all.
We discussed the reasons for this and concluded that the RFMCMC sampler acts as a pre-processor for our decoding algorithm, converting the uncorrectable state into a correctable state and allowing our decoding algorithm to decode to an error-free state. 
As a result, our decoding algorithm can eliminate errors very efficiently if the weight parameters are properly adjusted.
Our algorithm is deterministic and favorable because the decoding complexity is very low.
The current research also implies the correctness of the PP's insight that a combination of sophisticated classical decoding strategies and QA can solve problems that cannot be solved by either one alone.
We hope this study will contribute to the design of QA devices using SLHZ systems with near-term technology.

\begin{acknowledgments}
I would like to thank Dr. T. Kadowaki and Prof. H. Nishimori at the Tokyo Institute of Technology for their useful comments and discussions. 
I also thank Dr. Masayuki Shirane of NEC Corporation/National Institute of Advanced Industrial Science and Technology for his continuous support. 
This paper is partly based on results obtained from a project, JPNP16007, commissioned by the New Energy and Industrial Technology Development Organization (NEDO), Japan.
\end{acknowledgments}

\nocite{*}
\bibliography{APS}

\end{document}